\documentclass[a4paper,10pt]{article}
\usepackage[utf8]{inputenc}

\usepackage{graphicx} 
\usepackage{amsmath}
\usepackage{textcomp}
\usepackage{amsfonts}
\usepackage{multicol}
\usepackage{cite}
\usepackage{color}
\usepackage{pifont}
\usepackage{caption}
\usepackage{subcaption}
\usepackage{rotating}
\usepackage{enumerate}
\usepackage{bbm}

\usepackage{hyperref}

\normalbaselines
\title{Coherent states for the supersymmetric partners of the truncated oscillator}
\author{David J. Fern\'andez C.$^{1a}$, V\'eronique Hussin$^{2b}$, VS Morales-Salgado$^{1c}$
 \\ {\footnotesize $^1\,$ Departamento de F\'{\i}sica, Cinvestav,  A.P. 14-740, 07000 Ciudad de M\'exico, Mexico}
\\ {\footnotesize $^2\,$ D\'epartement de Math\'ematiques et de Statistique, Universit\'e de Montr\'eal, Montr\'eal, Qu\'ebec, H3C 3J7, Canada }
\\ {\footnotesize $^a\,$david@fis.cinvestav.mx, $^b\,$veronique.hussin@umontreal.ca, $^c\,$vmorales@fis.cinvestav.mx
}}
\begin{document}
\maketitle
\begin{abstract}
 We build the coherent states for a family of solvable singular Schr\"odinger Hamiltonians 
 obtained through supersymmetric quantum mechanics from the truncated oscillator.
 The main feature of such systems is the fact that their eigenfunctions
 are not completely connected by their natural ladder operators.
 We find a definition that behaves appropriately in the complete 
 Hilbert space of the system, through linearised ladder operators. 
 In doing so, we study basic properties of such states like 
 continuity in the complex parameter, resolution of the identity, 
 probability density, time evolution and possibility of entanglement.
\end{abstract}
\section{Introduction}
When studying quantum systems few tools have proven to be as fruitful 
as those describing them through coherent states (CS).
These states were originally introduced to characterise the most classical 
behaviour for the harmonic oscillator, despite the system being quantum in nature \cite{s26}.
Since then they became a widespread tool in the understanding of quantum phenomena
\cite{k63a,k63b,g63a,g63b,p86,gk99,aag00,q01,ah08}.

On the other hand, supersymmetric quantum mechanics is a spectral design method that allows 
obtaining solvable Hamiltonians departing from an initial one whose solution is already known.
Such a technique has been used to produce plenty of new solvable quantum systems
\cite{wi81,wi82,mi84,ais93,aicd95,cks95,bs97,fgn98,fhm98,jr98,qv99,sa99,mnr00,crf01,
ast01,mr04,cfnn04,ff05,cf08,ma09,f10,qe11,bf11a,ma12,ai12,ggm13}.
Although most of the work has focused on non-singular potentials, 
recent efforts start to consider potentials with singularities in their domains \cite{mnn98,fgn11}.
In particular, the truncated oscillator, 
a system described by a harmonic oscillator potential with an infinite barrier at the origin,
has shown to be an appropriate and simple model of this kind \cite{fm14,fm16,fm18}.
Moreover, there is a natural definition of ladder operators 
for the supersymmetric partners of the truncated oscillator, 
even though the energy spectra of such systems are divided into subsets, that cannot be connected through the action of these operators.
In fact, the spectral manipulation achievable for such a system is 
highly affected by the boundary condition at the origin.

Previously, it has been possible to define CS for the supersymmetric partners of non-singular potentials \cite{fhr07,bcf14}.
Thus, it is natural to proceed further and study such states for some singular cases.
However, earlier works have shown that a reduction process on the ladder operators is required in order to obtain appropriate CS for a kind of energy spectrum that some supersymmetric partners possess.
Linearised coherent states have been found to be a successful definition in these problematic cases \cite{bcf14}.
This was done for non-singular systems connected with the Painlev\'e IV equation.
In this work we shall extend the study as well to singular systems, that can be connected with the Painlev\'e V equation \cite{fm16,bfn16}.

This work is devoted to study several alternative definitions of coherent states 
for the truncated oscillator and its supersymmetric partners. 
In order to achieve this goal, the article is organised as follows:
In Section 2 we introduce the definitions of coherent states we are interested in.
In Section 3 we study the coherent states for the truncated oscillator, while in 
Section 4 we analyse these states for its supersymmetric partners. 
Finally, in Section 5 we present our final remarks and conclusions.
\section{Coherent states}\label{GCS}
In this section we will introduce in general terms several definitions of coherent states (CS) 
starting from a given set of operators $\{H,l^+,l^-\}$ characterising the system under study, 
where $H$ is a Schr\"odinger Hamiltonian and $l^\pm$ are its ladder operators,
that are not necessarily hermitian conjugate to each other.
We suppose that these operators can be realised in a separable Hilbert space $\mathcal{H}$.
In the Fock representation, an element of this basis is given by $|k\rangle$, 
where $k\in\mathbb{N}_0=\{0,1,2,...,d-1\}$, where $d$ is either finite or infinite,
and the action of these operators on a state $|k\rangle$ is given by:
\begin{eqnarray}\label{xik}
 H|k\rangle&=&\xi(k)\,|k\rangle\,,\\ \label{lm} 
 l^-|k\rangle&=&\sqrt{f(k)}\,\,|k-1\rangle\,,\\ \label{lp}
 l^+|k-1\rangle&=&\sqrt{g(k)}\,\,|k\rangle\,,
\end{eqnarray}
where $\xi(k)$ is a growing function of $k$ that defines 
the energy eigenvalue associated to $|k\rangle$ 
and $f(k)$ and $g(k)$ are functions of $k$ such
that $f(0)=0$ and, if $d$ is finite, $g(d)=0$.
\subsection{$l^-$ coherent states}\label{lcs}
The first definition of coherent states $|z\rangle$ to be considered, 
the $l^-$-CS, is given by the condition
\begin{equation}\label{acs}
 l^- \left| z \right\rangle =z\left|z\right\rangle\,,
\end{equation}
where $z\in \mathbb{C}$, i.e., $|z\rangle$ is an eigenstate 
of $l^-$ with complex eigenvalue $z$. 
Through a standard procedure \cite{p86,aag00}, $|z\rangle$ 
can be formally expanded in the Fock basis as
\begin{equation}
 |z\rangle=C_z\sum\limits_{k=0}^{d-1}\frac{z^k}{\sqrt{[f(k)]!}}|k\rangle\,,
\end{equation}
where $C_z=\left[\sum\limits_{k=0}^{d-1}\frac{|z|^{2k}}{[f(k)]!}\right]^{-1/2}$
is a normalisation constant, and 
\begin{equation}
 [f(k)]!=\left\{\begin{matrix}1\qquad&\text{for }&k=0\\
                        \prod\limits_{k'=1}^k\,f(k')&\text{for }&k>0\end{matrix}\right.
\end{equation}
is a generalisation of the factorial function.
Normalisability of these states is attained as long as $C_z$ remains finite.
For finite $d$ this fulfillment is immediate, but for an infinite dimensional Hilbert space it depends on the behavior of the function $f(k)$ when $k\to\infty$.

Notice that if the space $\mathcal{H}$ is finite-dimensional (dim$(\mathcal{H})=d$ 
with $d$ finite), the operator $l^-$ has a $d\times d$-matrix 
representation that is nilpotent of index $d$. 
Thus, the only solution to the exact eigenvalue equation (\ref{acs}) is $z=0$ \cite{fhr07,bcf14}.
Such a result makes the condition (\ref{acs}) unsuccessful for defining 
families of CS in a finite dimensional space $\mathcal{H}$. 
Work has been done in order to avoid this problem for some physical models with such a finite energy spectrum (like the Morse potential \cite{ah08}) by considering ``almost eigenstates of $l^-$'' .
\subsection{$D_l(z)$ coherent states}
An alternative definition of coherent states, the $D_l(z)$-CS, 
is given by the action of the so-called displacement operator $D_l(z)$
on an extremal state of the system, that we shall identify with $|0\rangle$.
The extremal states of a quantum system are those non-trivial eigenstates of $H$ that are in the kernel of the annihilation operator $l^-$.

Recall that the displacement operator for the harmonic oscillator is 
\begin{equation}
 D_a(z)=\,\text{exp}(za^+-z^*a^-)
       =\text{exp}\left(-\frac{1}{2}|z|^2\right)\text{exp}(za^+)\text{exp}(-z^*a^-)\,,
\end{equation}
where the second equation (the factorised expression) 
is derived from the commutation relation between $a^-$ and $a^+$.
For systems described by more general algebraic structures,
we use the following definition for the displacement operator 
(possibly non-unitary) \cite{p86}
\begin{equation}\label{DO}
 D_l(z)=\tilde C_z\,\text{exp}(zl^+)\text{exp}(-z^*l^-).
\end{equation}
It means that for the systems of interest in this work, 
the $D_l(z)$-CS are formally given by:
\begin{eqnarray}
 |z\rangle=D_l(z)|0\rangle
          =\tilde C_z\,\sum\limits_{k=0}^{d-1}\frac{\sqrt{[g(k)]!}}{k!}\,z^k\,|k\rangle.
\end{eqnarray}
The normalisation condition requires again that
$\tilde C_z=\left[\sum\limits_{k=0}^{d-1}\frac{[g(k)]!}{k!^2}|z|^{2k}\right]^{-1/2}$ 
which, as mentioned before, relies on its finiteness.
If $\mathcal{H}$ is infinite-dimensional, this cannot be guaranteed in general, since the main quantity to be controlled for obtaining an appropriate normalisation is the ratio
$ \frac{[g(k)]!}{k!^2} $ and its behavior for $k\to\infty$.
\subsection{Linearised coherent states}\label{RCS}
In order to reconciliate both definitions of CS ($l^-$-CS and $D_l(z)$-CS) it has been 
proposed to use a deformed version of the ladder operators $l^\pm$ \cite{fhr07} that linearises their commutator by using the associated number operator. 
Recently such a reduction was used to obtain a definition for 
$D_l(z)$-CS working well in both finite- and infinite-dimensional cases \cite{bcf14}. 
One of the reasons to look for a definition of coherent states that 
works well for both types of Hilbert spaces is to be able to generate them 
for systems as those that will be described in the following sections.
Indeed, let us introduce the linearised ladder operators $\ell^\pm$ 
through their action on the Fock basis:
\begin{eqnarray}\nonumber
 \ell^-\left|k\right\rangle=\sqrt{\alpha k}\,\,\left| k-1\right\rangle\,,\qquad
 \ell^+\left|k-1\right\rangle=\sqrt{\alpha k}\,\,\left|k\right\rangle\,.
\end{eqnarray}

In the case of an infinite-dimensional Hilbert space the generalised factorial
becomes $[f(k)]!=k!\,\alpha^k$ and thus the $\ell^-$-CS are given by:
\begin{equation}\label{lin-lcs}
 |z\rangle=C_z\sum\limits_{k=0}^{\infty}
           \frac{1}{\sqrt{k!}}
            \left(\frac{z}{\sqrt{\alpha}}\right)^k|k\rangle,
\end{equation}
where $C_z=\text{exp}\left(-\frac{|z|^2}{2\alpha}\right)$.
As indicated in section \ref{lcs}, in the case of a finite dimensional Hilbert space
such a definition of $\ell^-$-CS gives only the state $|0\rangle$.

On the other hand, having linearised the ladder operators 
makes the normalisability of the $D_{\ell}(z)$-CS straightforward. 
In fact, the linearised $D_{\ell}(z)$-CS are given by
\begin{equation}\label{lin-dcs}
 |z\rangle=\tilde C_z\,\sum_{k=0}^{d-1}
                 \frac{1}{\sqrt{k!}}\,(\sqrt{\alpha}z)^k\,|k\rangle\,,
\end{equation}
where 
$\tilde C_z=\left[\text{exp}\left(\alpha\,|z|^2\right)
     -\frac{|z|^{2d} \alpha^d}{\Gamma(d+1)}
      \,_2F_2\left(1,d+1;d+1,d+1;\alpha|z|^2\right)\right]^{-1/2}$
and $d={\rm dim}(\mathcal{H})$. In the infinite-dimensional case we have that
$\tilde  C_z=\text{exp}\left(-\alpha\frac{\,|z|^2}{2}\right)$. 
Note that, throughout this article
$_pF_q\left(a_1,...,a_p;b_1,...,b_q;x\right)$ represents the generalised hypergeometric function.

From equations (\ref{lin-lcs}) and (\ref{lin-dcs}) we can see that when the Hilbert space
is infinite dimensional ($d=\infty$) and $\alpha=1$ both definitions coincide 
and the standard expression for the CS for the harmonic oscillator is reproduced.
\section{Truncated oscillator and coherent states}
Let us remember that a quantum harmonic oscillator truncated by an infinite barrier at the origin is described by the Schr\"odinger Hamiltonian 
$H_0=-\frac{1}{2}\frac{{\rm d}^2}{{\rm d}x^2}+V_0(x)$, where
\begin{equation}
  V_{0}(x) = \left\{\begin{array}{l l}
                        \frac{x^{2}}{2} & \quad \text{if} \,\, x > 0        \\
                                 \infty & \quad \text{if} \,\, x \leq 0  \\  \end{array} \right., 
\end{equation}
is the potential of the system. 

The general solution to the stationary Schr\"odinger equation $H_0u(x)=\epsilon u(x)$ 
for $x>0$ is a linear combination of two solutions of definite parity, an odd and even one,
given by:
\begin{equation}\label{seedsol} 
  u(x,\epsilon)=e^{-x^2/2} \left[\,_1 F_1\left(\frac{1-2\epsilon}{4};\frac{1}{2};x^2\right)+2\nu \frac{\Gamma(\frac{3-2\epsilon}{4})}{\Gamma(\frac{1-2\epsilon}{4})}\,x\,_1 F_1\left(\frac{3-2\epsilon}{4};\frac{3}{2};x^2\right)\right]\,.
\end{equation}
Moreover, the boundary conditions of the problem, $\psi_k(0)=\psi_k(\infty)=0$, 
make the admissible eigenfunctions of $H_0$ to be 
\begin{equation}\label{1.0i} 
 \psi_k(x)=A_k\,x\,{\rm e}^{-x^2/2}\,_1\mbox{F}_1\left(-k;3/2;x^2\right)
          =B_k\,{\rm e}^{-x^2/2}{\rm H}_{2k+1}(x)\,,
\end{equation}
where $A_k=\left[\sqrt{\pi } 4^{k-1} (k!)^2/(2k+1)!\right]^{-1/2}$, 
$B_k=\left[\sqrt{\pi}\,4^k (2k+1)!\right]^{-1/2}$ 
and ${\rm H}_n$ is the $n$-th Hermite polynomial,
with corresponding eigenvalues $E_k=2k+\frac{3}{2}$, $k=0,1,...$
Note that the states $\psi_k$ of the truncated oscillator are related 
to the states $\psi_n^{\rm HO}$ of the harmonic oscillator restricted 
to the domain $(0,\infty)$ by $\psi_k=\sqrt{2}\,\psi_{2k+1}^{\rm HO}$.

The system described by $H_0$ has the natural ladder operators $l^\pm=(a^\pm)^2$, 
where $a^\pm$ are the ladder operators for the standard harmonic oscillator.
We thus get the commutation relations: 
\begin{eqnarray}
 \left[H_0,l^\pm\right]=\pm\,2l^\pm\,,\qquad \left[l^-,l^+\right]=4H_0\,.
\end{eqnarray}
Their action on a Fock state $|k\rangle$ associated to the eigenfunction
$\psi_k(x)=\langle x|k\rangle$ of equation (\ref{1.0i}) is given by 
\begin{eqnarray}
 l^-|k\rangle=\sqrt{2k(2k+1)}\,|k-1\rangle\,,\qquad\, 
 l^+|k-1\rangle=\sqrt{2k(2k+1)}\,|k\rangle\,.
\end{eqnarray}
It means that for this system we have $f(k)=g(k)=2k(2k+1)\,$.
\subsection{Coherent states and their properties}\label{CSg}
Such a system has an infinite dimensional Hilbert space $\mathcal{H}_0$,
and we examine next the three definitions of CS presented in section \ref{GCS}.

The $l^-$-CS are found as
\begin{equation}\label{tolcs}
 |z\rangle=C_z\sum\limits_{k=0}^\infty\,\frac{z^k}{\sqrt{(2k+1)!}}\,|k\rangle,
\end{equation}
where $C_z=\left[\frac{\sinh(|z|)}{|z|}\right]^{-1/2}\,$.

The $D_l(z)$-CS become:
\begin{eqnarray}\label{todcs}
 |z\rangle=\tilde C_z\,\sum\limits_{k=0}^\infty\,\frac{\sqrt{(2k+1)!}}{k!}\,z^k\,|k\rangle\,.
\end{eqnarray}
Such states are normalisable only for $|z|<1/2$, where the normalisation constant 
is $\tilde C_z=(1-4|z|^2)^{3/4}$, making this definition a poor choice.
The third choice leads to the linearised CS already given in 
equations (\ref{lin-lcs}) and (\ref{lin-dcs}) when $\alpha=2$.

Let us study next some properties of the $l^-$-CS.
First of all, through a standard procedure \cite{bcf14}, we can show that these coherent states are continuous in the complex parameter $z$.
Even more, they provide a resolution of the identity in $\mathcal{H}$, 
$\mathbbm{1}=\int|z\rangle\langle z|\,\mu(z)\,dz$, where the integral measure is
\begin{equation}
 \mu(z)=\frac{|z|^2}{8\,\pi C_z^2}\,{\rm exp}(-|z|)\,.
\end{equation}

\begin{figure}[h]
 \centering
 \includegraphics[width=0.6\textwidth]{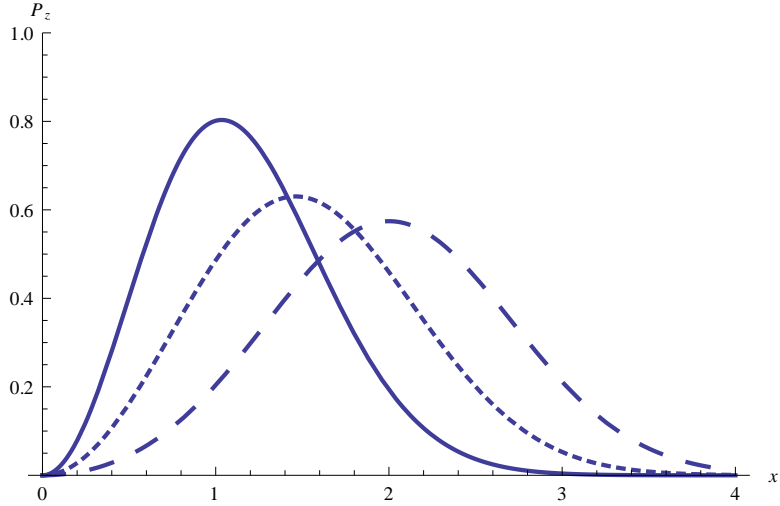}
 \caption{$P_z(x)$ for the $l^-$-CS at $|z|=0.1$ ({\bf \textendash\textemdash\textemdash}), 
          $|z|=1$ (- - - -), $|z|=2$ ($-\,\,-\,\,-$).}\label{PA}
\end{figure}

The probability density distribution for a $l^-$-CS $|z\rangle$ 
is given by $P_z(x)=|\langle x|z\rangle|^2$, where
\begin{equation}\label{ProbDens}
 \langle x|z\rangle=\sum_{n=0}^{\infty}\,\langle x|n\rangle\langle n|z\rangle
                   =C_z\,\sum_{n=0}^{\infty}\frac{z^n}{\sqrt{(2n+1)!}}\,\psi_n(x)\,.
\end{equation}       
Plots for some particular examples of such a probability density can be found in figure \ref{PA}.

We could also compute the state probability, i.e., the probability 
$p_n(z)=|\langle n|z\rangle|^2 $ of obtaining the eigenvalue $\xi(n)$ 
as a result of an energy measurement when the system is in the coherent state 
$|z\rangle$ (see equation (\ref{xik})), that is given by
\begin{equation}
 p_n(z)=C^2_z\,\frac{|z|^{2n}}{(2n+1)!}\,.
\end{equation}

The time evolution of a $l^-$-CS is obtained through the expression
\begin{eqnarray}
 U(t)|z\rangle=\text{exp}(-iH_0t)|z\rangle\,.
\end{eqnarray}
Since the energy eigenvalues $E_k$ are linear in $k$,
these coherent states are temporally stable, i.e., 
an arbitrary $l^-$-CS evolves into another state that is also a $l^-$-CS. 

It is worth mentioning that these properties were analysed also for the $\ell^-$-CS 
and the results do not pose great difference from those of the $l^-$-CS. 
The linearised coherent states are continuous in $z$ and temporally stable.
They provide a resolution of the identity and their probability density
shows well localisation similar to that of the $l^-$-CS depicted in figure \ref{PA}.
\subsection{Measures of entanglement}    
Now, we shall turn to the question of whether these CS can be used to produce entanglement.
To deal with this possibility let us study two measures of entanglement:
the uncertainty relation and the linear entropy.
For this we need to compute explicitly expectation values in the coherent states, obtained here as follows. 

For an observable $O$ it is possible to obtain its expectation value in a $l^-$-CS
in the Schr\"odinger picture through the equation
\begin{eqnarray}\label{evo}
 \langle O\rangle=\langle z|O|z\rangle
                 =\sum_{m,n=0}^{\infty}\,\Lambda_{mn}(z)\,\langle O_{nm} \rangle\,,
\end{eqnarray}
where 
$\Lambda_{mn}(z)=C_z^2\,\frac{z^m(z^*)^n}{\sqrt{(2m+1)!(2n+1)!}}$ 
and $\langle O_{nm} \rangle=\int\,\psi^*_{n}(x)\,O\,\psi_m(x)\,{\rm d}x\,$.
In order to simplify future calculations let us use the fact that $O$ is hermitian
to rewrite equation (\ref{evo}) as follows
\begin{equation}
 \langle O\rangle=\sum_{n=0}^\infty\,\Lambda_{nn}(z)\,\langle O_{nn} \rangle 
                  +\sum_{n=1}^\infty\sum_{m=0}^{n-1}\,2\,{\rm Re}\left(\Lambda_{mn}(z)\,\langle O_{nm}\rangle\right) .
\end{equation}
For example, if the observable $O$ is chosen as the Hamiltonian $H_0$,
we obtain the expectation value of the energy $\langle H_0\rangle=\frac{1}{2}+|z|\coth\left(|z|\right)$.

Before moving on, let us recall that on the domain $(0,\infty)$ the momentum operator $p=-i\frac{\rm d}{{\rm d}x}$ is not an observable,
as can be asserted by the fact that its deficiency indices are $(1,0)$, and thus, by the deficiency theorem $p$ has no self-adjoint extensions \cite{W10,vN29}.
Nonetheless, we are still interested in studying the behavior of the expectation value of the operator $p$.

\subsubsection{Uncertainty relation.}
We can now compute the standard deviations of the position and momentum operators
$\sigma_x=\sqrt{\langle x^2\rangle-\langle x\rangle^2}$,
$\sigma_p=\sqrt{\langle p^2\rangle-\langle p\rangle^2}$, 
and ultimately their product $\sigma_x\sigma_p\,$.
The lower bound of this product characterises the uncertainty relation 
for the pair of operators $x$ and $p$.
Let us recall that, since the ladder operators are now $l^\pm=(a^\pm)^2$,
then $x$ and $p$ cannot be written in a simple manner in terms of $l^\pm$.

In order to compute the uncertainty relation for the operators $x$ and $p$, 
we use the matrix elements 

{\footnotesize
\begin{eqnarray}\nonumber
 \langle x_{nm}\rangle&=&
   \sqrt{\frac{(2n+1)!}{(2m+1)!}}\frac{(-2)^{n-m-1}\Gamma(n-m-\frac{1}{2})}{\pi\,(2n-2m)!}\\&&\qquad\qquad\times\,_2F_1\left[-2m-1,n-m-\frac{1}{2};2(n-m)+1;2\right]\,,\\
 \langle x^2_{nm}\rangle&=&
   \frac{1}{2}\delta_{nm}+\frac{\sqrt{\,(2n+1)!(2m+1)!}}{(n+m)!}\left(\frac{1}{2}\delta_{n,m-1}+\delta_{n,m}+\frac{1}{2}\delta_{n,m+1}\right)\,,\\\nonumber\\\nonumber 
 \langle p_{nm}\rangle&=&
   i\,\langle x_{nm}\rangle-\frac{i(2m+1)}{\pi}\,\sqrt{\frac{(2n+1)!}{(2m+1)!}}\frac{(-2)^{n-m}(2n-2m-1)\Gamma\left(n-m-\frac{1}{2}\right)}{(2n-2m+1)!}\\&&\qquad\qquad\times\,_2F_1\left[n-m+\frac{1}{2},-2m;2(n-m+1);2\right]\,,\\\nonumber
 \langle p^2_{nm}\rangle&=&
   \frac{1}{2}\delta_{nm}-2\sqrt{2m(2m+1)}\delta_{m-1,n}\\&&\qquad\qquad\,+\frac{2\sqrt{(2n+1)!(2m+1)!}}{(n+m)!}\left(\frac{3}{2}\delta_{n,m-1}+\delta_{n,m}-\frac{1}{2}\delta_{n,m+1}\right)\,,\\\nonumber   
\end{eqnarray} }
where $n\geq m$.

\begin{figure}[h]
 \centering
 \includegraphics[width=0.6\textwidth]{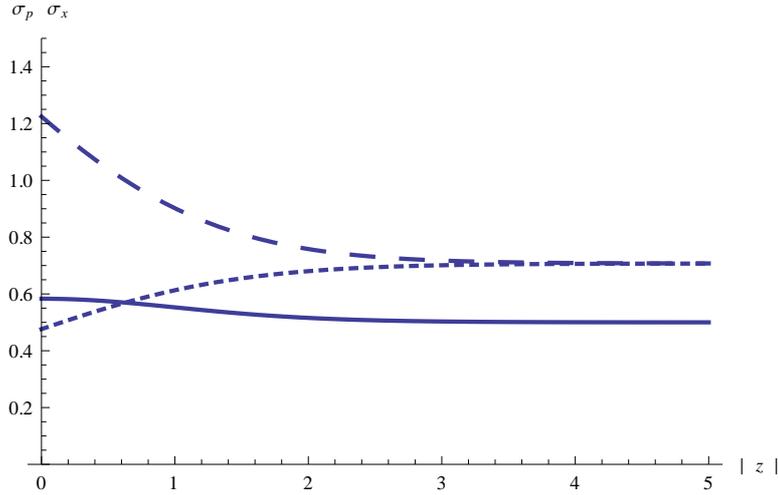} 
 \caption{Uncertainty relation for the $l^-$-CS: 
          $\sigma_p$ ($-\,\,-\,\,-$), $\sigma_x$ (- - - -), 
          $\sigma_x\sigma_p$ ({\bf \textendash\textemdash\textemdash}).}\label{UncA}
\end{figure}

Figure \ref{UncA} shows approximations of the corresponding standard deviations 
and their product as functions of the modulus of the complex parameter $z$. 
These were obtained by truncating the infinite sums in equation (\ref{evo}) up to the 30th term.
We can see that the dispersion of the position $x$ 
is always lower than that of the momentum $p$. 
However, as $|z|$ increases both $\sigma_x$ and $\sigma_p$ approach the value
$1/\sqrt{2}$ and thus their product tends to the value $1/2$.

\begin{figure}[h]
 \centering
 \includegraphics[width=0.6\textwidth]{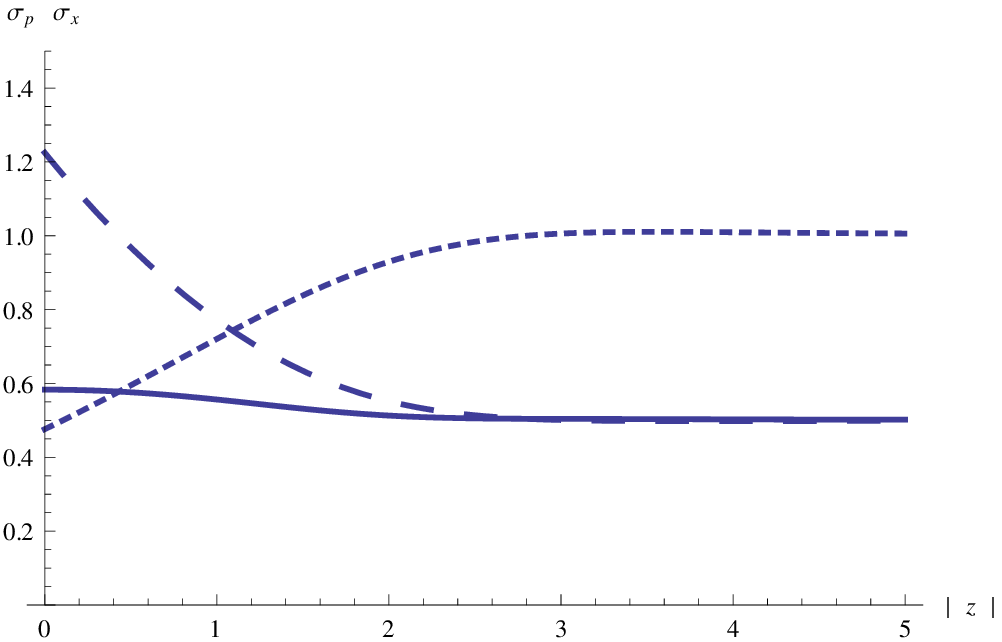} 
 \caption{Uncertainty relation for the $\ell^-$-CS: 
          $\sigma_p$ ($-\,\,-\,\,-$), $\sigma_x$ (- - - -), 
          $\sigma_x\sigma_p$ ({\bf \textendash\textemdash\textemdash}).}\label{rUncA}
\end{figure}

For the linearised coherent states, figure \ref{rUncA} 
shows the results of the approximations carried out up to the 30th term of (\ref{evo}).
Observe that $\sigma_p$ is a decreasing function of $|z|$ while $\sigma_x$ is
a growing function of $|z|$.
For $0<|z|<1$ the dispersion of the momentum is greater than that of the position
and they become equal for $|z|=1$. 
For $1<|z|$ the dispersion of the position surpasses that of the momentum.
Opposite to the results for the $l^-$-CS, 
in this case we can see that the dispersions of 
$x$ and $p$ do not approach to each other as $|z|$ increases.
However, once again the product $\sigma_x\sigma_p$ tends to $1/2$ as $|z|$ increases.

In any case, these graphs yield an uncertainty relation such that
\begin{equation}
 \sigma_x\sigma_p\geq\frac{1}{2}\,,
\end{equation}
and, since $\sigma_x$ and $\sigma_p$ are different from each other 
in the case of linearised as well as non-linearised coherent states,
we can say that there is squeezing phenomena \cite{w83,lk87}, that suggests the possibility of producing entangled states by means of these coherent states.
\subsubsection{Entanglement and linear entropy.}\label{enlinen}
As a measure of entanglement let us use the beam splitter tool 
and compute the linear entropy of the out-state \cite{fl87,twc91,sw01,ksbk02,x02}.
By definition the beam splitter operator is 
$\mathcal{B}={\rm e}^{\frac{\theta}{2}(a^+b\,{\rm e}^{i\varphi}-ab^+\,{\rm e}^{-i\varphi})}$,
where $\theta$ is the angle of the beam splitter and $\varphi$ 
is the phase difference between the reflected and transmitted states,
from which the reflection and transmission amplitudes are
$r=-{\rm e}^{-i\varphi}\sin(\theta/2)$, $t=\cos(\theta/2)$, respectively.

The beam splitter acts on an input that is a bipartite state 
$|\rm in\rangle$ yielding the out-state \cite{gk05}:
\begin{equation}\label{outstate}
 |{\rm out}\rangle=\mathcal{B}|{\rm in}\rangle
                  ={\rm e}^{\frac{\theta}{2}(a^+b\,{\rm e}^{i\varphi}-ab^+\,{\rm e}^{-i\varphi})}|{\rm in}\rangle\,,
\end{equation}
where the in-state is such that $|{\rm in}\rangle\in\mathcal{H}_1\otimes\mathcal{H}_2$, $a$, $a^+$ and $b$, $b^+$ are the standard first-order ladder operators 
of the harmonic oscillator in each space of the tensor product.
Notice that the operators $K_-=ab^+$, $K_+=a^+b$, $K_0=\frac{1}{2}(a^+a-b^+b)$ 
generate a $su(2)$ algebra.

In our case, $|{\rm in}\rangle=|0\rangle\otimes| z \rangle$
where $| z \rangle$ is a $l^-$-CS. We thus get
\begin{equation}
 |{\rm out}\rangle=\mathcal{C}_z\sum_{n}\,\frac{1}{\sqrt{(2n+1)!}}\,z^n\,\mathcal{B}(|0,n\rangle)\,.
\end{equation}

Using this last state we can define the density operator 
$\rho_{AB}=|{\rm out}\rangle \langle{\rm out}|$, 
and by taking a partial trace we obtain $\rho_A$. 
The linear entropy is thus $S=1-{\rm Tr}(\rho_A^2)$, 
and we know that $S=0$ corresponds to no entanglement 
while $S=1$ corresponds to maximal entanglement.
           
Let us recall that $|n\rangle=\sqrt{2}\,|2n+1\rangle_{\rm HO}$; thus
\begin{flalign}\nonumber
 &\qquad\quad\mathcal{B}|0,n\rangle
 =2\,{\rm e}^{\frac{\theta}{2}(a^+b\,{\rm e}^{i\varphi}-ab^+\,{\rm e}^{-i\varphi})}|1,2n+1\rangle_{\rm HO}&\\\nonumber
 &\qquad\quad\qquad\,\,\,=2\,\sum_{k=0}^{2n+2}\frac{1}{k!}\left[{\rm e}^{i\varphi}\tan\frac{\theta}{2}\right]^k\left[\left(\cos\frac{\theta}{2}\right)^{2n}\sqrt{\frac{(k+1)!(2n+1)!}{(2n+1-k)!}}\,|k+1,2n+1-k\rangle_{\rm HO}\right.&\\
 &\qquad\qquad\qquad\quad\left.-\tan(\theta/2)\,{\rm e}^{-i\varphi}\sqrt{2n+2}\left(\cos\frac{\theta}{2}\right)^{(2n+2)}\sqrt{\frac{(k)!(2n+2)!}{(2n+2-k)!}}\,|k,2n+2-k\rangle_{\rm HO} \right]&
\end{flalign}
where we have used the Baker-Campbell-Haussdorf formula for the $su(2)$ algebra 
\cite{t85} generated by $K_-=ab^+$, $K_+=a^+b$, $K_0=\frac{1}{2}(a^+a-b^+b)$: 
\begin{equation}
 {\rm e}^{\tau K_+-\tau^*K_-}={\rm e}^{\frac{\tau}{|\tau|}\tan|\tau|K_+}
                              {\rm e}^{-2(\ln\cos|\tau|)K_0}
                              {\rm e}^{-\frac{\tau^*}{|\tau|}\tan|\tau|K_-}\,.
\end{equation}

In terms of the reflection and transmission amplitudes we obtain
\begin{eqnarray}\nonumber
 |{\rm out}\rangle&=&\mathcal{C}_z\sum_{n=0}^\infty\,\frac{z^n}{\sqrt{(2n+1)!}}\,
  \sum_{k=0}^{2n+2}\frac{\left(-r/t\right)^k}{k!}\,{\rm e}^{i2k\varphi}\,\\
  &&\qquad\qquad\times\left[Q_1\,|k+1,2n+1-k\rangle_{\rm HO}+Q_2\,|k,2n+2-k\rangle_{\rm HO} \right]\,,
\end{eqnarray}
where
\begin{eqnarray}\nonumber
 Q_1(r,t,n,k)&=&t^{2n}\sqrt{\frac{(k+1)!(2n+1)!}{(2n+1-k)!}}\,,\\\nonumber
 Q_2(r,t,n,k)&=&r\,t^{(2n+1)}\sqrt{2n+2}\,\sqrt{\frac{(k)!(2n+2)!}{(2n+2-k)!}}\,.
\end{eqnarray}
     
Remember that the bra-ket product is obtained by an integration over the domain $(0,\infty)$. 
In fact \cite{pbm90}
\begin{eqnarray}\nonumber
 (\langle\alpha|\beta\rangle)_{\rm HO}=\int_0^\infty{\rm e}^{-x^2}H_\alpha(x)H_\beta(x)\,dx
  =\sqrt{\pi}\,\frac{\,_2F_1\left[-\alpha,-\beta;1-\frac{\alpha+\beta}{2};1/2\right]}{2^{1-\alpha-\beta}\Gamma\left(1-\frac{\alpha+\beta}{2}\right)}\,.
\end{eqnarray}

Now the density operator $\rho_{AB}=|{\rm out}\rangle \langle{\rm out}|$ is given by
\begin{flalign}\nonumber
 &\qquad\quad\rho_{AB}
 =\mathcal{C}_z^2\sum_{m,n=0}^\infty\,\frac{z^n(z^*)^m}{\sqrt{(2n+1)!(2m+1)!}}\,\sum_{k=0}^{2n+2}\sum_{l=0}^{2m+2}\frac{(-r/t)^k}{k!}\frac{(-r^*/t)^l}{l!}\,{\rm e}^{i2(k-l)\varphi}&\\\nonumber
  &\qquad\qquad\qquad\times\left[Q^*_2(r,t,m,l)Q_2(r,t,n,k)\,(|k,2n+2-k\rangle\langle l,2m+2-l|)_{\rm HO}\right.&\\\nonumber
  &\qquad\qquad\qquad\qquad+\,Q^*_1(r,t,m,l)Q_2(r,t,n,k)\,(|k,2n+2-k\rangle\langle l+1,2m+1-l|)_{\rm HO}&\\\nonumber
  &\qquad\qquad\qquad\qquad+\,Q^*_2(r,t,m,l)Q_1(r,t,n,k)\,(|k+1,2n+1-k\rangle\langle l,2m+2-l|)_{\rm HO}&\\\nonumber
  &\qquad\qquad\qquad\qquad\left.+\,Q^*_1(r,t,m,l)Q_1(r,t,n,k)\,(|k+1,2n+1-k\rangle\langle l+1,2m+1-l|)_{\rm HO}\right]\,.&
\end{flalign}
     
\begin{figure}[h]
 \centering
 \includegraphics[width=.6\linewidth]{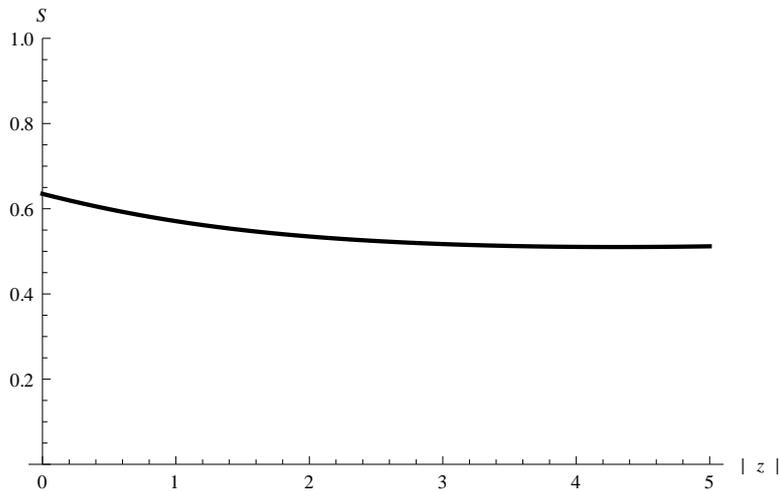}
 \caption{Linear entropy for the $l^-$-CS with beam splitter angles 
          $\theta=\pi/2$, $\varphi=0.$}\label{csS}
\end{figure}

Computing now the partial trace $\rho_A$ we get the linear entropy $S=1-{\rm Tr}(\rho_A^2)$.
Figure \ref{csS} shows numerical approximations for $S(|z|)$,
obtained by truncating the resulting infinite sum up to the 20th term.

It is worth noticing that the plot is overall flat and it shows that 
the out-sate is entangled, although not maximally entangled.
If we repeat the computations now for the linearised coherent states,
the plot of $S(|z|)$ behaves in general in the same way as with the $l^-$-CS.
\section{The supersymmetric truncated oscillator and coherent states}
In quantum mechanics, a supersymmetric transformation relates 
two Hamiltonians as follows \cite{cks95,jr98,f10,mnn98}. 
Suppose that the Hamiltonian $H=-\frac{1}{2}\frac{{\rm d}^2}{{\rm d}x^2}+V(x)$ 
is obtained through a $q$-th order supersymmetric transformation from an 
initial Hamiltonian $H_0=-\frac{1}{2}\frac{{\rm d}^2}{{\rm d}x^2}+V_0(x)$, 
i.e., there exists a differential operator $Q$ of order $q$, 
called {\it intertwining operator}, such that the following relation holds
\begin{equation}
 HQ=QH_0\,.
\end{equation}

Up to an overall shift in the energy, the set of eigenvalues of $H$ differs 
in a finite number, $\kappa=\kappa(q)$, of elements from that of $H_0$. 
This means that the supersymmetric transformation has added and/or erased 
levels in the energy spectrum of $H_0$ to obtain the one of $H$.

If the energy spectrum of $H_0$ is infinite countable, then the Hilbert space $\mathcal{H}$ 
spanned by the eigenfunctions of $H$ is the direct sum of two subspaces: 
$\mathcal{H}=\mathcal{H}_{\rm iso}\oplus\mathcal{H}_{\rm new}$,
an infinite-dimensional one denoted by $\mathcal{H}_{\rm iso}$
and a $\kappa$-dimensional subspace denoted by $\mathcal{H}_{\rm new}$.
The eigenvectors of $H$ are given by the union of those 
$|E_n\rangle\in\mathcal{H}_{\rm iso}$ and the ones
$|\mathcal{E}_j\rangle\in\mathcal{H}_{\rm new}$.
Then, the energy spectrum of $H$ is also given by a union of two sets.
The first one is an infinite set given by $\{E_0,E_1,E_2,\cdots\}$ while the second one is the finite set $\{\mathcal{E}_0,\mathcal{E}_1,\cdots,\mathcal{E}_{\kappa-1}\}$.

If the initial system admits ladder operators $l^\pm$, 
then the system described by $H$ has natural ladder operators in 
the subspace $\mathcal{H}_{\rm iso}$ given by the product $Q\,l^\pm\,Q^\dagger$, 
where $Q^\dagger$ is the hermitian adjoint of $Q$.
On the other hand, in $\mathcal{H}_{\rm new}$ such a definition does not yield 
in general operators that connect the eigenenergies in this subspace. 
However, in the following section we will use a reduction theorem that will 
allow us to obtain well-behaved ladder operators in $\mathcal{H}_{\rm new}$.
\subsection{Supersymmetric partners of the truncated oscillator}\label{susytho}
The supersymmetric partners of the truncated oscillator \cite{fm14,fm16,fm18},
obtained through a $q$-th order supersymmetric transformation, 
are described by Hamiltonians $H=-\frac{1}{2}\frac{{\rm d}^2}{{\rm d}x^2}+V$ 
with potentials given by \cite{cks95,jr98,f10,mnn98}
\begin{equation}
 V=V_0-\{\text{ln}\left[W\left(u_1,...,u_q\right)\right]\}'' ,
\end{equation}
where $W\left(u_1,...,u_q\right)$ is the Wronskian of $q$ {\it seed solutions} 
of the initial Schr\"odinger equation $H_0 u_j=\epsilon_j u_j$, 
$j=1,...,q$, given in general by (\ref{seedsol}).
These $u_j$'s do not need to satisfy any boundary conditions, 
although they must not introduce new singularities to the superpartner potential $V$.

Without loss of generality, we can order the $\epsilon_j$'s as $\epsilon_1<...<\epsilon_q$.
Moreover, as is usual we shall suppose that $\epsilon_q<\frac{1}{2}$.
Under these assumptions, the set of eigenvalues of $H$ has 
$\kappa=\left[\frac{q}{2}\right]$ (the integer part of $\frac{q}{2}$) 
new elements corresponding to the eigenvalues $\mathcal{E}_i$ 
added by the non-singular supersymmetric transformation, 
which associated eigenfunctions indeed satisfy the boundary conditions \cite{fm18}.
Hence, the eigenvalues in $\mathcal{H}_{\rm iso}$ form an infinite ladder 
$\{E_0,E_1,E_2,...\}$ such that $E_k=E_0+2k$, where $E_0=3/2$, 
while in $\mathcal{H}_{\rm new}$ the eigenvalues 
$\{\mathcal{E}_0,\mathcal{E}_1,...,\mathcal{E}_{\kappa-1}\}$
are not in general equally spaced.

Now, in order to define ladder operators working appropriately on both subspaces,
we have to consider that the eigenvalues in $\mathcal{H}_{\rm new}$ 
are related by $\mathcal{E}_j=\mathcal{E}_0+2j$. 
Thus, the partial spectrum $\{\mathcal{E}_0,\mathcal{E}_1,...,\mathcal{E}_{\kappa-1}\}$ forms 
an equidistant set of eigenvalues, and by means of a reduction theorem 
applied to the operators $Q\,(a^\pm)^2\,Q^\dagger$,
where $a^\pm$ are the standard first-order ladder operators of the harmonic oscillator,
it is possible to obtain now sixth-order ladder operators $L^\pm$ in the whole space $\mathcal{H}$ that satisfy \cite{bf11a,bcf14}
\begin{equation}\label{num}
 L^+L^-=\left(H-1/2\right)\left(H-3/2\right)
        \left(H-\epsilon_1\right)\left(H-\epsilon_2\right)
        \left(H-\epsilon_{q-1}-2 \right)\left(H-\epsilon_q-2 \right).
\end{equation}
Moreover, the commutation relations $\left[H,L^\pm \right]=\pm 2L^\pm$ still hold,
and they yield the following action of $L^\pm$ on the basis vectors of $\mathcal{H}$.

In $\mathcal{H}_{\rm iso}$:
\begin{eqnarray}\nonumber
 &L^-\left|E_n\right\rangle
  =\left[\left(E_n-1/2\right)\left(E_n-3/2\right)
         \left(E_n-\epsilon_1\right)\left(E_n-\epsilon_2 \right)\right.&\\\label{l-iso}
 &\qquad\qquad\qquad\times\left.\left(E_n-\epsilon_{q-1}-2\right)\left(E_n-\epsilon_q-2 \right)\right]^{1/2}
    \,\,\left| E_{n-1}\right\rangle\,,&\\ \nonumber
 &L^+\left|E_n\right\rangle
  =\left[\left(E_{n+1}-1/2\right)\left(E_{n+1}-3/2\right)
          \left(E_{n+1}-\epsilon_1\right)\left(E_{n+1}-\epsilon_2\right)\right.&\\\label{l+iso} 
 &\qquad\qquad\qquad\times\left.\left(E_{n+1}-\epsilon_{q-1}-2\right)\left(E_{n+1}-\epsilon_q-2 \right)\right]^{1/2}
    \,\,\left| E_{n+1}\right\rangle\,.&
\end{eqnarray} 
In $\mathcal{H}_{\rm new}$:
\begin{eqnarray}\nonumber
 &L^-\left|\mathcal{E}_j\right\rangle
  =\left[\left(\mathcal{E}_j-1/2\right)\left(\mathcal{E}_j-3/2\right)
         \left(\mathcal{E}_j-\epsilon_1\right)\left(\mathcal{E}_j-\epsilon_2 \right)\right.\\ \label{l-new}
  &\qquad\qquad\left.\times\left(\mathcal{E}_j-\epsilon_{q-1}-2\right)\left(\mathcal{E}_j-\epsilon_q-2 \right)\right]^{1/2}
   \,\,\left|\mathcal{E}_{j-1}\right\rangle\,,\\ \nonumber
 &L^+\left|\mathcal{E}_j\right\rangle
  =\left[\left(\mathcal{E}_{j+1}-1/2\right)\left(\mathcal{E}_{j+1}-3/2\right)
         \left(\mathcal{E}_{j+1}-\epsilon_1\right)\left(\mathcal{E}_{j+1}-\epsilon_2 \right)\right.\\ \label{l+new}
  &\qquad\qquad\left.\times\left(\mathcal{E}_{j+1}-\epsilon_{q-1}-2\right)\left(\mathcal{E}_{j+1}-\epsilon_q-2 \right)\right]^{1/2}
   \,\,\left|\mathcal{E}_{j+1}\right\rangle\,. 
\end{eqnarray} 
From these results we can see that $L^-$ annihilates the eigenstates 
corresponding to $\frac{3}{2}$ and $\mathcal{E}_0$, while $L^+$ 
annihilates the eigenstate corresponding to $\mathcal{E}_{\kappa-1}$.

It is well known that the differential operators $L^-$, $L^+$ and $H$ defining second and third degree polynomial Heisenberg algebras are determined by solutions of the 
Painlev\'e IV and Painlev\'e V equations respectively
\cite{s92,vs93,a94,dek94,ekk94,srk97,acin00,mn08}.
In \cite{fm14,fm16} it was explicitly shown how different supersymmetric partners of the truncated oscillator, 
ruled by appropriate $L^-$, $L^+$ and $H$, are thus connected to 
these non-linear second-order differential equations.
In \cite{bcf14} coherent states were built for systems connected to the Painlev\'e IV equation. 
Here we will build coherent states for supersymmetric partners of the truncated oscillator that are ruled by both, the second and fifth degree polynomial Heisenberg algebras.

When trying to define coherent states in $\mathcal{H}$ using $L^\pm$, one realises that 
these ladder operators do not connect energy levels from 
$\mathcal{H}_{\rm iso}$ with those in $\mathcal{H}_{\rm new}$, 
which justifies the direct sum decomposition of $\mathcal{H}$. 
Thus, we require a definition of coherent states appropriate 
for finite- and infinite-dimensional spaces simultaneously. 

From section \ref{GCS} we know that the $L^-$-CS are appropriate for 
$\mathcal{H}_{\rm iso}$ but fail  in $\mathcal{H}_{\rm new}$. 
The opposite happens with $D_L(z)$ coherent states:
while this definition is appropriate for $\mathcal{H}_{\rm new}$ 
it fails in general in $\mathcal{H}_{\rm iso}$.
Therefore, we will implement the definition of linearised coherent states from section \ref{RCS} for the supersymmetric partners of the truncated oscillator.
As previously remarked, each subspace, $\mathcal{H}_{\rm iso}$ or $\mathcal{H}_{\rm new}$, 
possesses one extremal state given by its lowest energy state,
thus allowing one to apply the displacement operator in each case.
Before doing this, however, we will illustrate with an example how
the supersymmetry method applied to the truncated oscillator works.

Let us consider the case where a fourth order ($q=4$) 
supersymmetric transformation has been implemented,
with $\epsilon_1=-\frac{11}{2}$, $\epsilon_2=-\frac{9}{2}$, 
$\epsilon_3=-\frac{7}{2}$, $\epsilon_4=-\frac{5}{2}$.
Then we get that $\kappa=2$, and the eigenstates in $\mathcal{H}_{\rm new}$ 
are those associated to the eigenvalues $\mathcal{E}_0=\epsilon_2=-\frac{9}{2}$
and $\mathcal{E}_1=\epsilon_4=-\frac{5}{2}$.
The supersymmetric partner potential thus obtained is given by

{\scriptsize
\begin{equation}\label{susypotential}
 V(x)=\frac{x^2}{2}-\frac{4\left(256 x^{16}-2560 x^{14}+10496 x^{12}-19584 x^{10}+27360 x^8-10080 x^6-10800 x^4+16200 x^2+2025\right)}{\left(16 x^8-64 x^6+120 x^4+45\right)^{2}} 
\end{equation}  }
where $x\in(0,\infty)$.

The subspace isospectral to the truncated oscillator $\mathcal{H}_{\rm iso}$ is spanned by
\begin{eqnarray}\label{phin}
 \phi_n =\frac{Q\psi_n}{\sqrt{\prod\limits_{i=1}^{4}(E_n-\epsilon_i)}},
\end{eqnarray}
where $\psi_n$ is given by equation (\ref{1.0i}) and the intertwining operator 
\begin{equation}
 Q=\frac{1}{4}\left(\frac{{\rm d}^4}{{\rm d}x^4}+\eta_3(x)\frac{{\rm d}^3}{{\rm d}x^3}
   +\eta_2(x)\frac{{\rm d}^2}{{\rm d}x^2}+\eta_1(x)\frac{{\rm d}}{{\rm d}x}+\eta_0(x)\right)
\end{equation}
is a fourth-order differential operator whose coefficients are given by
\begin{eqnarray}
 \eta_3&=&-\frac{4 x \left(16 x^8-32 x^6+24 x^4+120 x^2+45\right)}{16 x^8-64 x^6+120 x^4+45},\\
 \eta_2&=&\frac{6 \left(16 x^{10}-16 x^8+88 x^6-120 x^4+165 x^2-105\right)}{16 x^8-64 x^6+120 x^4+45},\\
 \eta_1&=&\frac{4 x \left(-16 x^{10}+16 x^8-216 x^6+48 x^4+915 x^2+315\right)}{16 x^8-64 x^6+120 x^4+45},\\
 \eta_0&=&\frac{16 x^{12}-32 x^{10}+360 x^8+240 x^6-795 x^4-7110 x^2+1935}{16 x^8-64 x^6+120 x^4+45}.
\end{eqnarray}
The lowering operator $L^-$ acts on the basis of this subspace as follows
\begin{equation}
 L^-\left|E_n\right\rangle=\sqrt{\left(E_n-1/2\right)\left(E_n-3/2\right)\left(E_n+11/2\right)\left(E_n+9/2\right)\left(E_n+3/2 \right)\left(E_n+1/2 \right)}\,\,\left| E_{n-1}\right\rangle.
\end{equation}

On the other hand, the only two eigenfunctions in $\mathcal{H}_{\rm new}$ are given by
\begin{eqnarray}\label{phivarepsilon}
 \phi_{\mathcal{E}_0}&=&\frac{4 \sqrt{3} e^{-\frac{x^2}{2}} x \left(8 x^6-4 x^4+10 x^2+15\right)}{\sqrt[4]{\pi } \left(8 \left(2 x^4-8 x^2+15\right) x^4+45\right)},\\
 \phi_{\mathcal{E}_1}&=&\frac{2 e^{-\frac{x^2}{2}} x \left(16 x^8+72 x^4-135\right)}{\sqrt{3} \sqrt[4]{\pi } \left(8 \left(2 x^4-8 x^2+15\right) x^4+45\right)},
\end{eqnarray}
associated to $\mathcal{E}_0=-9/2$, $\mathcal{E}_1=-5/2$, respectively.
The lowering operator $L^-$ acts on the corresponding basis vectors as follows
\begin{equation}
 L^-\left|\mathcal{E}_j\right\rangle
  =\sqrt{\left(\mathcal{E}_j-1/2\right)\left(\mathcal{E}_j-3/2\right)\left(\mathcal{E}_j+11/2\right)\left(\mathcal{E}_j+9/2\right)\left(\mathcal{E}_j+3/2\right)\left(\mathcal{E}_j+1/2\right)}\,\,\left|\mathcal{E}_{j-1}\right\rangle.
\end{equation}
\subsection{Linearised $D_{\mathcal{L}}(z)$ coherent states}
A linearisation of $L^\pm$ can be obtained by defining 
the following linearised ladder operators
\begin{eqnarray}\nonumber
  \mathcal{L}^+:= \gamma(H) L^+\,,  \qquad  \mathcal{L}^-:= \gamma(H+2) L^-\,, 
\end{eqnarray}
where $\gamma(H)=\left[(H-1/2)(H-\epsilon_1)(H-\epsilon_2)(H-\epsilon_{q-1}-2)(H-\epsilon_{q}-2)\right]^{-1/2}$.
 
The action of $\mathcal{L}^\pm$ on the basis vectors of $\mathcal{H}_{\rm iso}$ is
\begin{eqnarray}\label{ell-iso}
 \mathcal{L}^-\left|E_n\right\rangle=\sqrt{E_n-3/2}\,\,\left| E_{n-1}\right\rangle=\sqrt{2n}\,\,\left| E_{n-1}\right\rangle,\\ \label{ell+iso}
 \mathcal{L}^+\left|E_n\right\rangle=\sqrt{E_{n+1}-3/2}\,\,\left| E_{n+1}\right\rangle=\sqrt{2n+2}\,\,\left| E_{n+1}\right\rangle.
\end{eqnarray}
Thus, the operator $\mathcal{L}^-$ annihilates the state $|E_0\rangle$, 
which is the extremal state in $\mathcal{H}_{\rm iso}$.
Also, note that the commutators between $\mathcal{L}^\pm$ and $H$ 
are those of a Heisenberg-Weyl algebra in this subspace
\begin{equation}\label{hwalg}
 \left[H,\mathcal{L}^\pm\right]=\pm2\mathcal{L}^\pm, \qquad \left[\mathcal{L}^-,\mathcal{L}^+\right]=2.
\end{equation}

Acting the displacement operator
$D_{\mathcal{L}}(z)=\text{exp}\left(-|z|^2\right)\text{exp}\left(z\mathcal{L}^+\right)\text{exp}\left(-z^*\mathcal{L}^-\right)$
on the extremal state $|E_0\rangle$ we obtain the desired linearised CS in $\mathcal{H}_{\text{iso}}$:
\begin{eqnarray}\nonumber
 |z\rangle_{\text{iso}}=D_{\mathcal{L}}(z)|E_0\rangle
  =\text{exp}\left(-|z|^2\right)\sum_{n=0}^{\infty}\frac{(\sqrt{2}z)^n}{\sqrt{n!}}|E_n\rangle. 
\end{eqnarray}

On the other hand, the action of $\mathcal{L}^\pm$ on 
the basis vectors of $\mathcal{H}_{\text{new}}$ is given by
\begin{eqnarray}\label{ell-new}
 \mathcal{L}^-\left|\mathcal{E}_j\right\rangle&=&(1-\delta_{j,0})\sqrt{\mathcal{E}_j-3/2}\,\,|\mathcal{E}_{j-1}\rangle
   =(1-\delta_{j,0})\sqrt{2j-\Delta_1}\,\,|\mathcal{E}_{j-1}\rangle,\\ \label{ell+new}
 \mathcal{L}^+\left|\mathcal{E}_j\right\rangle&=&(1-\delta_{j,\kappa-1})\sqrt{\mathcal{E}_{j+1}-3/2}\,\,|\mathcal{E}_{j+1}\rangle
   =(1-\delta_{j,\kappa-1})\sqrt{2j+2-\Delta_1}\,\,|\mathcal{E}_{j+1}\rangle\,,
\end{eqnarray}
where $\Delta_1=\frac{3}{2}-\mathcal{E}_0$.
Note that $\mathcal{L}^-$ annihilates the state $|\mathcal{E}_0\rangle$,
while $\mathcal{L}^+$ annihilates the state $|\mathcal{E}_{\kappa-1}\rangle$ so that $|\mathcal{E}_0\rangle$ is the extremal state in $\mathcal{H}_{\rm new}$.
Although one needs to be cautious when these operators act on
$|\mathcal{E}_0\rangle$ and $|\mathcal{E}_{\kappa-1}\rangle$, 
in general, the commutators in (\ref{hwalg}) still hold when acting on the other basis vectors of $\mathcal{H}_{\rm new}$.

Displacing now the state $|\mathcal{E}_0\rangle$ gives us 
the following CS in $\mathcal{H}_{\text{new}}$:
\begin{eqnarray}\nonumber
 |z\rangle_{\text{new}}=D_{\mathcal{L}}(z)|\mathcal{E}_0\rangle
  =\hat C_z\,\sum_{j=0}^{\kappa-1}\frac{(\sqrt{2}z)^j}{j!}
         \sqrt{\frac{\Gamma(-\frac{\Delta_1}{2}\,+j)}{\Gamma(-\frac{\Delta_1}{2})}}\,|\mathcal{E}_{j}\rangle\,,
\end{eqnarray}
where
\begin{eqnarray}\nonumber
 \hat C_z=\, _1F_1\left[-\frac{\Delta_1}{2};1;2|z|^2\right]
     -\frac{|z|^{2\kappa} 2^\kappa \Gamma \left(\frac{2\kappa-\Delta_1}{2}\right)}
           {\left[\Gamma(\kappa+1)\right]^2 \Gamma \left(-\frac{\Delta_1}{2}\right)}
      \,_2F_2\left[1,\kappa-\frac{\Delta_1}{2};\kappa+1,\kappa+1;2|z|^2\right] .
\end{eqnarray}

The $D_{\mathcal{L}}(z)$-CS are continuous in the complex parameter $z$ 
and they lead to a resolution of the identity, where the integral measures in each subspace $\mathcal{H}_{\rm iso}$ and $\mathcal{H}_{\rm new}$ are given respectively by
\begin{equation}\label{mu1}
 \mu_{\rm iso}(z)=\frac{2}{\pi}\,, \qquad
 \mu_{\rm new}(z)=\frac{2\,\Gamma\left(\frac{-\Delta_1}{2}\right)}{\pi \hat C_z^2}
         \,G^{2,0}_{1,2}
          \left(\begin{matrix}-\left(\frac{\Delta_1+2}{2}\right)\\0\,,\,0\end{matrix}\,
           \bigg|\,2|z|^2 \right)\,,          
\end{equation}
where $G^{m,n}_{p,q}\left(\begin{matrix}a_1,...,a_p\\b_1,...,b_q\end{matrix}\,      \bigg|\,x\right)$ is the Meijer G-function. 
As in the non-supersymmetric case, these states are temporally stable.

The mean value of the energy in each subspace reads
\begin{equation}
 \langle H\rangle_{\rm iso}=\frac{3}{2}+4|z|^2\,,\qquad
 \langle H\rangle_{\rm new}=
  \mathcal{E}_0+2|\hat C_z|^2\sum_{j=0}^{\sigma-1}\frac{j(2\,|z|^2)^j}{(j!)^2\Gamma\left(\frac{\Delta_1}{2}-j\right)}\,,
\end{equation}
while the state probability is given by
\begin{eqnarray}\nonumber
 P_n(z)=\frac{(2|z|^2)^n}{n!}\text{exp}\left(-2|z|^2\right)\,,\qquad
 P_j(z)=\frac{(2\,|z|^2)^{j-1}}{\left[(j-1)!\right]^2}\frac{|\hat C_z|^2}{\Gamma\left(\frac{\Delta_1}{2}+1-j\right)}\,,
\end{eqnarray}
in $\mathcal{H}_{\rm iso}$ and $\mathcal{H}_{\rm new}$, respectively.

\begin{figure}[h]
  \includegraphics[width=0.48\textwidth]{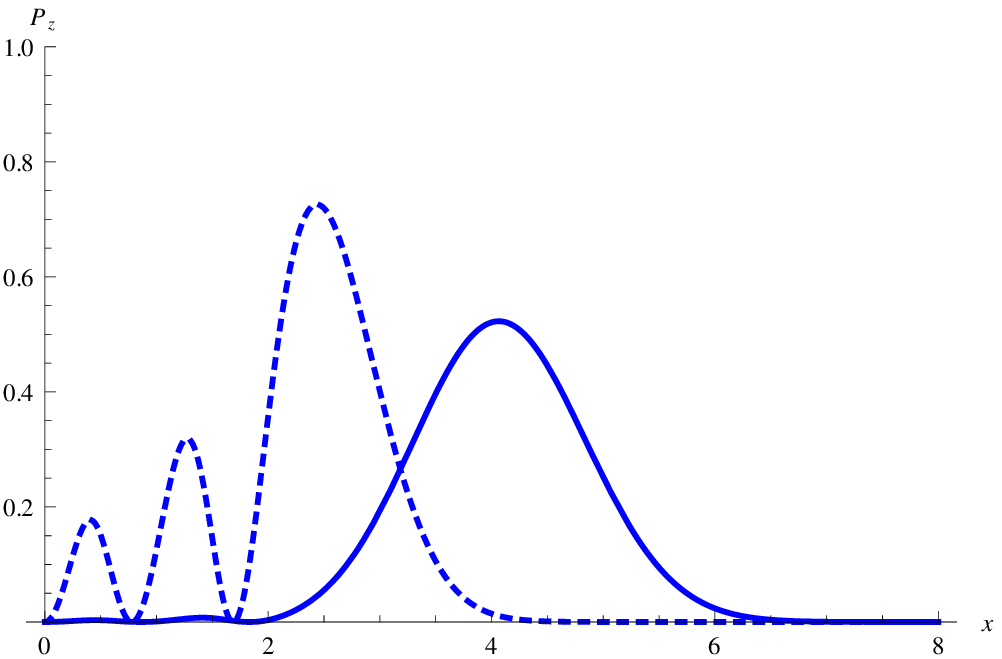}
  \hspace*{\fill}
  \includegraphics[width=0.48\textwidth]{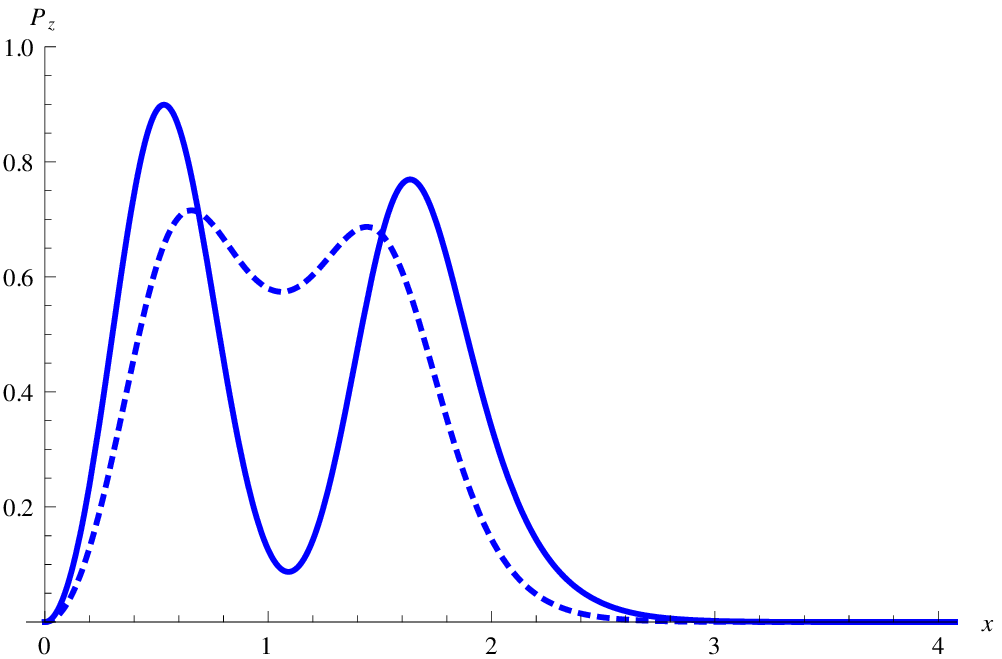}\\ 
  \hspace*{\fill}$\mathcal{H}_{\rm iso}$\hspace*{\fill}\hspace*{\fill}$\mathcal{H}_{\rm new}$\hspace*{\fill}
 \caption{Probability density of the coherent states for $|z|=0.1$ (- - -) 
          and $|z|=10$ ({\bf \textemdash\textemdash}).}\label{ssPz}
\end{figure}

Consider again the previous example, 
where the levels $\mathcal{E}_0=-9/2$ and $\mathcal{E}_0=-5/2$ 
have been added to the energy spectrum of the supersymmetric partner.
The probability density $P(x)=|\langle x|z\rangle|^2$ in each subspace of 
$\mathcal{H}$ is depicted in figure \ref{ssPz}, showing fairly good localisation.
We notice that for $|z\rangle_{\rm iso}$ this localisation 
is somewhat lost for small values of $|z|$. 
In the case of $|z\rangle_{\rm new}$, 
for big values of $|z|$ two distinct peaks occur,
that however remain close to each other. 
Greater localisation is achieved for small values of $|z|$ since the peaks start to merge.
\subsection{Measures of entanglement}  
Following the results and notation in equation (\ref{evo}) we obtain the matrix elements
\begin{eqnarray}\nonumber
 \Lambda_{m,n}(z)&=&{\rm exp}(-2|z|^2)\,\frac{(\sqrt{2}z)^m(\sqrt{2}z^*)^{n}}{\sqrt{m!n!}}\,,\\ \nonumber
 \Lambda_{i,j}(z)&=&\hat C_z^2\,\frac{(\sqrt{2}z)^i(\sqrt{2}z^*)^{j}}{i!j!}\,
  \frac{\sqrt{\Gamma\left(-\frac{\Delta_1}{2}+i\right)\Gamma\left(-\frac{\Delta_1}{2}+j\right)}}
       {\Gamma\left(-\frac{\Delta_1}{2}\right)}\,,
\end{eqnarray}
in $\mathcal{H}_{\rm iso}$ and $\mathcal{H}_{\rm new}$, respectively.
\subsubsection{Uncertainty relation.}
\begin{figure}[h]
 \includegraphics[width=0.48\textwidth]{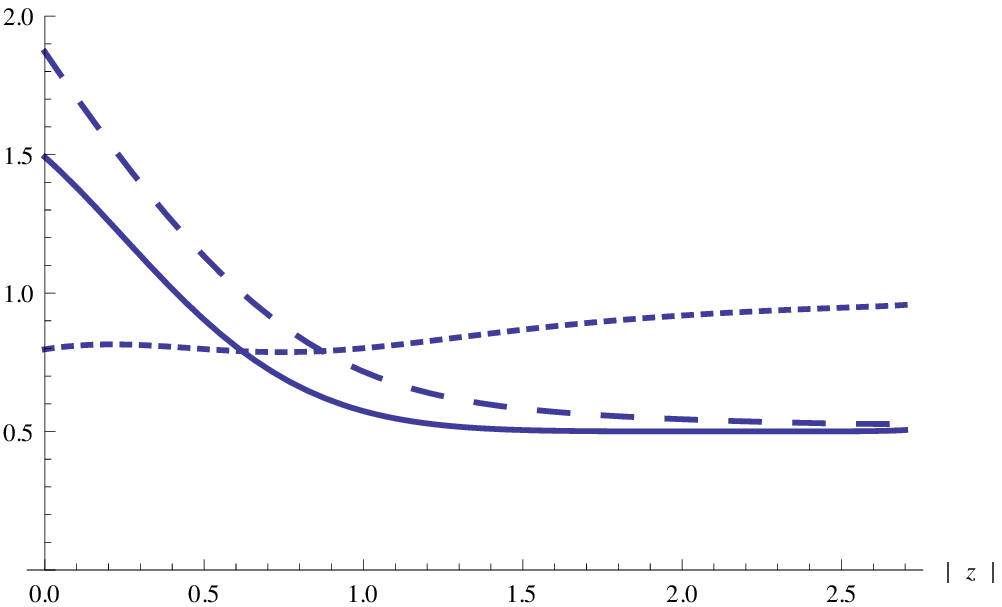}
 \hspace*{\fill}
 \includegraphics[width=0.48\textwidth]{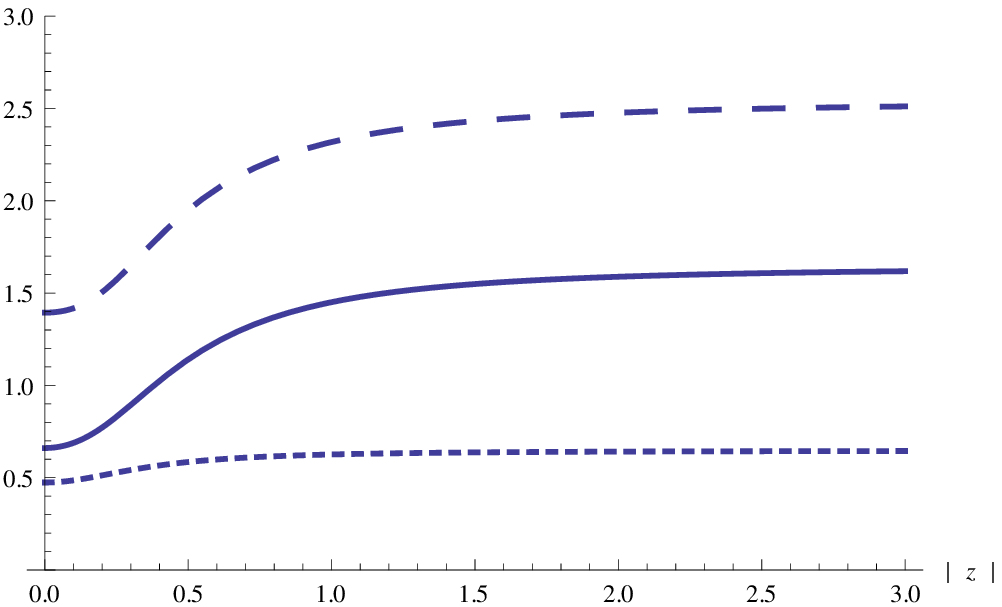}\\ 
  \hspace*{\fill}$\mathcal{H}_{\rm iso}$\hspace*{\fill}\hspace*{\fill}$\mathcal{H}_{\rm new}$\hspace*{\fill}
 \caption{Uncertainty relation for the $D_{\mathcal{L}}(z)$-CS:
          $\sigma_x$ (- - - -), $\sigma_p$ ($-\,\,-\,\,-$), 
          $\sigma_x\sigma_p$ ({\bf \textendash\textemdash\textemdash}).}\label{ssDU}
\end{figure}

Let us rely on the example introduced at the end of section \ref{susytho}, where a fourth 
order supersymmetric transformation was carried out for the truncated oscillator.
Recall that in this example $\mathcal{H}_{\rm new}$ was two-dimensional,
since the added eigenfunctions spanning it were those associated to 
the eigenvalues $\mathcal{E}_0=-9/2$ and $\mathcal{E}_1=-5/2$, 
while the space $\mathcal{H}_{\rm iso}$ remained infinite-dimensional,
as it is isospectral to the truncated oscillator.
Figure \ref{ssDU} shows the behavior of $\sigma_x$, 
$\sigma_p$ and their product $\sigma_x\sigma_p$, 
as functions of the modulus of the complex parameter $z$.

In both subspaces we can observe squeezing phenomena:
while in $\mathcal{H}_{\rm new}$ the uncertainty in the momentum is always 
greater than the uncertainty in the position, in $\mathcal{H}_{\rm iso}$ we find
the squeezing of $\sigma_x$ for $|z|<1$ and the squeezing of $\sigma_p$ for $|z|>1$.
This indicates once again that entanglement is attainable by means of these coherent states.
\subsubsection{Entanglement and linear entropy.}
Let us obtain the linear entropy in each of the two subspaces 
$\mathcal{H}_{\rm iso}$ and $\mathcal{H}_{\rm new}$.

The in-states in these subspaces are given by 
\begin{equation}
 |{\rm in}\rangle_{\rm iso}={\rm e}^{-|z|^2}\,\sum_{i=0}^{\infty}\frac{(\sqrt{2}z)^i}{\sqrt{i!}}\,|\mathcal{E}_0,E_i\rangle\,,
\qquad
 |{\rm in}\rangle_{\rm new}=\hat C_z\,\sum_{j=0}^{\kappa-1}\frac{(\sqrt{2}z)^j}{j!}
  \sqrt{\frac{\Gamma(-\frac{\Delta_1}{2}\,+j)}{\Gamma(-\frac{\Delta_1}{2})}}\,|\mathcal{E}_0,\mathcal{E}_{j}\rangle\,,
\end{equation}
respectively.
By means of equation (\ref{outstate}) we can obtain the out-states
$|{\rm out}\rangle_{\rm iso}$, $|{\rm out}\rangle_{\rm new}$.
Through a similar treatment as in section \ref{enlinen}, for which $\varphi=0$, 
we find that, in coordinates representation the out-states are given by
\begin{flalign}\nonumber
 &\qquad\langle x,y|{\rm out}\rangle_{\rm iso}
  =\mathcal{\tilde C}_z\,\sum_{j=0}^{\infty} \frac{(\sqrt{2}z)^j}{\sqrt{j!}}
      \sum_{l,m,n=0}^\infty\frac{1}{l!m!n!}\left[\tan\frac{\theta}{2}\right]^n\left[\ln(\cos\frac{\theta}{2})\right]^m\left[-\,\tan\frac{\theta}{2}\right]^l&\\\nonumber
 &\qquad\qquad\qquad\qquad\times\sum_{i=0}^m\binom{m}{i}\,(-1)^{-i}\partial_x^n\,x^i\,\partial_x^i\left[x^l\,\langle x|\mathcal{E}_0\rangle\right]\,y^{m+n-i}\langle y|E_{j}\rangle^{(l+m-i)}\,,&          
\end{flalign}
\begin{flalign}\nonumber
 &\qquad\langle x,y|{\rm out}\rangle_{\rm new}
  =\mathcal{\hat C}_z\,\sum_{j=0}^{\kappa-1} \frac{(\sqrt{2}z)^j}{j!}\sqrt{\frac{\Gamma(-\frac{\Delta_1}{2}\,+j)}{\Gamma(-\frac{\Delta_1}{2})}}
      \sum_{l,m,n=0}^\infty\frac{1}{l!m!n!}\left[\tan\frac{\theta}{2}\right]^n\left[\ln(\cos\frac{\theta}{2})\right]^m\left[-\,\tan\frac{\theta}{2}\right]^l&\\\nonumber
 &\qquad\qquad\qquad\qquad\times\sum_{i=0}^m\binom{m}{i}\,(-1)^{-i}\partial_x^n\,x^i\,\partial_x^i\left[x^l\,\langle x|\mathcal{E}_0\rangle\right]\,y^{m+n-i}\langle y|\mathcal{E}_{j}\rangle^{(l+m-i)}\,,&
\end{flalign}
in $\mathcal{H}_{\rm iso}$ and $\mathcal{H}_{\rm new}$, respectively, where $\mathcal{\tilde C}_z$ and $\mathcal{\hat C}_z$ are normalisation constants.      
      
As before, we use the out-state to obtain the operator 
$\rho=|{\rm out}\rangle\langle{\rm out}|$
and then we compute the linear entropy $S=1-{\rm Tr}(\rho_A^2)$.

\begin{figure}[h]\centering
 \includegraphics[width=0.6\textwidth]{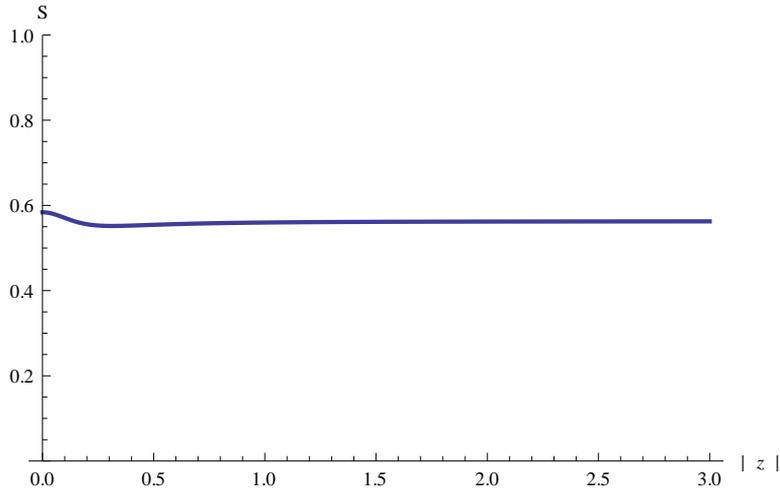}
  \caption{Approximation to the linear entropy $S$ for the angles $\theta=\pi/2$ and $\varphi=0$ 
           and real values of the complex parameter $z$.}\label{ssSnew}
\end{figure}

For the system described by the potential in equation (\ref{susypotential}),
a numerical approximation to the linear entropy in $\mathcal{H}_{\rm new}$ 
as function of $|z|$ can be found in figure \ref{ssSnew}. 
The infinite series were truncated at the 30th term.
Although being flat, with values close to $0.5$, 
this plot show evidence of entanglement, but not of a maximally entangled out-state.
In the case of the subspace $\mathcal{H}_{\rm iso}$ the behavior of the linear entropy 
is similar to that depicted in figure \ref{ssSnew}.
\section{Conclusions}
In this work we studied several definitions of coherent states 
for the truncated oscillator and its supersymmetric partners.
The main difficulty in defining coherent states for systems obtained through
a supersymmetric transformation lays in the fact that the resulting energy spectrum 
is not completely connected by the action of the natural ladder operators,
leading to a decomposition of $\mathcal{H}$ as a direct sum of two subspaces: 
a finite dimensional one and other infinite dimensional.

It has been found that the conventional CS definitions that work well for finite 
dimensional spaces fail when applied to infinite dimensional ones and vice versa.
We surpassed such a difficulty by using a reduced linearised version of the natural ladder operators for the system.
Thus we arrived to a common definition of coherent states, that can be applied to both subspaces.
Consequently, we studied some properties of these coherent states.
We found the production of squeezing in the position and momentum of these coherent states,
as well as the arising of entanglement in the out-state of a beam splitter whose in-state 
was one of these coherent states.

\section{Acknowledgments}
This work has been supported in part by research grants from 
Natural sciences and engineering research council of Canada (NSERC).
The authors acknowledge the financial support of the Spanish MINECO 
(project MTM2014-57129-C2-1-P) and Junta de Castilla y Le\'on (VA057U16).
VS Morales-Salgado also acknowledges the Conacyt fellowship 243374
and the Department of Foreign Affairs, Trade and Development of Canada
for the Emerging Leaders in the Americas Program (ELAP) scholarship provided.


\section*{References}
\begin{thebibliography}{100}
\bibitem{s26} Schr\"odinger E 1926 {\it Naturwiss} {\bf 14} 664
\bibitem{k63a} Klauder J R 1963 {\it J. Math. Phys.} {\bf 4} 1955
\bibitem{k63b} Klauder J R 1963 {\it J. Math. Phys.} {\bf 4} 1958
\bibitem{g63a} Glauber R J 1963 {\it Phys. Rev.} {\bf 130} 2529
\bibitem{g63b} Glauber R J 1963 {\it Phys. Rev.} {\bf 131} 2766
\bibitem{p86} Perelomov A 1986 {\it Generalized Coherent States and their Applications} (Springer) 
\bibitem{gk99} Gazeau J P and Klauder J R 1999 {\it J. Phys. A: Math. Gen.} {\bf 32} 123
\bibitem{aag00} Ali S T, Antoine J P and Gazeau J P 2014 {\it Coherent States, Wavelets and their Generalizations, 2nd Ed} (Springer)
\bibitem{q01} Quesne C 2001 {\it Ann. Phys.} {\bf 293} 147
\bibitem{ah08} Angelova M and Hussin V 2008 {\it J. Phys. A} {\bf 41} 30416

\bibitem{wi81} Witten E 1981 {\it Nucl. Phys. B} {\bf 185} 513
\bibitem{wi82} Witten E 1982 {\it Nucl. Phys. B} {\bf 202} 253
\bibitem{mi84} Mielnik B 1984 {\it J. Math. Phys.} {\bf 25} 3387
\bibitem{ais93} Andrianov A A, Ioffe M V and Spiridonov V P 1993 {\it Phys. Lett. A} {\bf 174} 273
\bibitem{aicd95} Andrianov A A, Ioffe M V, Cannata F and Dedonder J P 1995 {\it Int. J. Mod. Phys. A} {\bf 10} 2683
\bibitem{cks95} Cooper F, Khare A and Sukhatme U 1995 {\it Phys. Rep.} {\bf 251} 267
\bibitem{bs97} Bagrov V G and Samsonov B F 1997 {\it Phys. Part. Nucl.} {\bf 28} 374
\bibitem{fgn98} Fern\'andez D J, Glasser M L and Nieto L M 1998 {\it Phys. Lett. A} {\bf 240} 15
\bibitem{fhm98} Fern\'andez D J, Hussin V and Mielnik B 1998 {\it Phys. Lett. A} {\bf 244} 309
\bibitem{jr98} Junker G and Roy P {\it Ann. Phys.} {\bf 270} 155
\bibitem{qv99} Quesne C and Vansteenkiste N 1999 {\it Helv. Phys. Acta} {\bf 72} 71
\bibitem{sa99} Samsonov B F 1999 {\it Phys.Lett. A} {\bf 263} 274
\bibitem{mnr00} Mielnik B, Nieto L M and Rosas-Ortiz O 2000 {\it Phys. Lett. A} {\bf 269} 70
\bibitem{crf01} Cari\~nena J F, Ramos A and Fern\'andez D J 2001 {\it Ann. Phys.} {\bf 292} 42
\bibitem{ast01} Aoyama H, Sato M and Tanaka T 2001 {\it Nucl. Phys. B} {\bf 619} 105
\bibitem{mr04} Mielnik B and Rosas-Ortiz O, {\it J. Phys. A: Math. Gen.} {\bf 37} 10007
\bibitem{cfnn04} Carballo J M, Fern\'andez D J, Negro J and Nieto L M 2004 {\it J. Phys. A: Math. Gen.} {\bf 37} 10349
\bibitem{ff05} Fern\'andez D J and Fern\'andez-Garc\'ia N 2005 {\it AIP Conference Proceedings} {\bf 744} 236
\bibitem{cf08} Contreras-Astorga A and Fern\'andez D J 2008 {\it J. Phys. A: Math. Theor.} {\bf 41} 475303
\bibitem{ma09} Marquette I 2009 {\it J. Math. Phys.} {\bf 50} 095202
\bibitem{f10} Fern\'andez D J 2010 {\it AIP Conf. Proc.} {\bf 1287} 3
\bibitem{qe11} Quesne C 2011 {\it Mod. Phys. Lett. A} {\bf 26} 1843
\bibitem{bf11a} Berm\'udez D and Fern\'andez D J 2011 {\it SIGMA} {\bf 7} 025
\bibitem{ma12} Marquette I 2012 {\it J. Math. Phys.} {\bf 53} 012901
\bibitem{ai12} Andrianov A A and Ioffe M V 2012 {\it J. Phys. A: Math. Theor.} {\bf 45} 503001
\bibitem{ggm13} G\'omez-Ullate D, Grandati Y and Milson R 2014 {\it J. Phys. A: Math. Theor.} {\bf 47} 015203

\bibitem{mnn98} M\'arquez I F, Negro J and Nieto L M 1998 {\it J. Phys. A: Math. Gen.} {\bf 31} 4115
\bibitem{fgn11} Fern\'andez D J, Gadella M and Nieto L M 2011 {\it SIGMA} {\bf 7} 029
\bibitem{fm14} Fern\'andez D J and Morales-Salgado V S 2014 {\it J. Phys. A: Math. Theor.} {\bf 47} 035304
\bibitem{fm16} Fern\'andez D J and Morales-Salgado V S 2016 {\it J. Phys. A: Math. Theor.} {\bf 49} 195202
\bibitem{fm18} Fern\'andez D J and Morales-Salgado V S 2018 {\it Ann. Phys.} {\bf 388} 122

\bibitem{fhr07} Fern\'andez D J, Hussin V and Rosas-Ortiz O 2007 {\it J. Phys. A: Math. Theor.} {\bf 40} 6491 
\bibitem{bcf14} Bermudez D, Contreras-Astorga A and Fern\'andez D J 2014 {\it Ann. Phys.} {\bf 350} 615
\bibitem{bfn16} Bermudez D, Fern\'andez D J and Negro J 2016 {\it J. Phys. A: Math. Theor.} {\bf 49} 335203

\bibitem{s92} Shabat A 1992 {\it Inverse Problems} {\bf 8} 303
\bibitem{vs93} Veselov A P and Shabat A B 1993 {\it Funct. Anal. Appl.} {\bf 27} 81
\bibitem{a94} Adler V E 1994 {\it Physica D} {\bf 73} 335
\bibitem{dek94} Dubov S Y, Eleonskii V M and Kulagin N E 1994 {\it Chaos} {\bf 4} 47
\bibitem{ekk94} Eleonskii V M, Korolev V G and Kulagin N E 1994 {\it Chaos} {\bf 4} 583
\bibitem{srk97} Sukhatme U P, Rasinariu C and Khare A 1997 {\it Phys. Lett. A} {\bf 234} 401
\bibitem{acin00} Andrianov A A, Cannata F, Ioffe M and Nishnianidze D 2000 {\it Phys. Lett. A} {\bf 266} 341
\bibitem{mn08} Mateo J and Negro J 2008 {\it J. Phys. A: Math. Theor.} {\bf 41} 045204

\bibitem{W10} Weyl H 1910 {\it Math. Ann.} {\bf 68} 220
\bibitem{vN29} von Neumann J 1929 {\it Math. Ann.} {\bf 102} 49
\bibitem{t85} Truax D R 1985 {\it Phys. Rev. D} {\bf 31} 1988
\bibitem{pbm90} Prudnikov A P, Brychkov Y A and Marichev O I 1990 {\it Integrals and Series, Vol. 2: Special Functions} (Gordon and Breach)

\bibitem{w83} Walls D F 1983 {\it Nature} {\bf 306} 141 
\bibitem{lk87} Loudon R and Knight P L 1987 {\it J. Mod. Opt.} {\bf 34} 709

\bibitem{fl87} Fearn H and Loudon R 1987 {\it Opt. Commun.} {\bf 64} 485
\bibitem{twc91} Tan S M, Walls D F and Collett M J 1991 {\it Phys. Rev. Lett.} {\bf 66} 252
\bibitem{sw01} Scheel S and Welsch D G 2001 {\it Phys. Rev. A} {\bf 64} 063811 
\bibitem{ksbk02} Kim M S, Son A, Bužek V and Knight P L 2002 {\it Phys. Rev. A} {\bf 65} 032323 
\bibitem{x02} Xiang-Bin W 2002 {\it Phys. Rev. A} {\bf 66} 024303 

\bibitem{gk05} Gerry C and Knight P 2005 {\it Introductory Quantum Optics} (Cambridge University Press)

\end{thebibliography}
\end{document}